\documentclass[aps,prx,
reprint,
superscriptaddress,
showpacs,
longbibliography,
]{revtex4-1}
\usepackage{amsmath}
\usepackage{amssymb}
\usepackage{natbib}
\usepackage[squaren,Gray]{SIunits}
\usepackage{graphicx}
\usepackage{dcolumn}
\usepackage{bm}

\usepackage[hidelinks,colorlinks=true,linktoc=page,linkcolor=blue,citecolor=blue,urlcolor=blue]{hyperref}
\begin{document}

\preprint{APS/123-QED}

\title{Higgs-mode radiance and charge-density-wave order in 2H-NbSe$_2$}
\author{Romain Grasset}
\affiliation{Universit\'e Paris Diderot, Sorbonne Paris Cit\'e, CNRS Laboratoire Mat\'eriaux et Ph\'enom\`{e}nes Quantiques, UMR 7162 75013, Paris, France}
\author{Tommaso Cea}
\affiliation{IMDEA Nanoscience, C/Faraday 9, 28049 Madrid, Spain}
\affiliation{Graphene Labs, Fondazione Istituto Italiano di Tecnologia, Via Morego, 16163 Genova, Italy}
\affiliation{ISC-CNR and Department of Physics, Sapienza University of Rome, P.le A. Moro 5, 00185 Rome, Italy}
\author{Yann Gallais}
\author{Maximilien Cazayous}
\author{Alain Sacuto}
\affiliation{Universit\'e Paris Diderot, Sorbonne Paris Cit\'e, CNRS Laboratoire Mat\'eriaux et Ph\'enom\`{e}nes Quantiques, UMR 7162 75013, Paris, France}
\author{Laurent Cario}
\affiliation{Institut des Mat\'eriaux Jean Rouxel (IMN), Universit\'e de Nantes - CNRS, 2 rue de la Houssini\'ere, BP 32229, 44322 Nantes Cedex 03, France.}
\author{Lara Benfatto}
\email[author to whom correspondence should be addressed: ]{ lara.benfatto@roma1.infn.it}
\affiliation{ISC-CNR and Department of Physics, Sapienza University of Rome, P.le A. Moro 5, 00185 Rome, Italy}
\author{Marie-Aude M\'easson}
\email[author to whom correspondence should be addressed: ]{ marie-aude.measson@neel.cnrs.fr}
\email[now at : Institut N\'eel, CNRS and Universit\'e Grenoble Alpes, F-38042 Grenoble, France]{}
\affiliation{Universit\'e Paris Diderot, Sorbonne Paris Cit\'e, CNRS Laboratoire Mat\'eriaux et Ph\'enom\`{e}nes Quantiques, UMR 7162 75013, Paris, France}



%


\date{\today}

\begin{abstract}
Despite being usually considered two competing phenomena, charge-density-wave and superconductivity coexist in few systems, the most emblematic one being the transition metal dichalcogenide 2H-NbSe$_2$. This unusual condition is responsible for specific Raman signatures across the two phase transitions in this compound. While the appearance of a soft phonon mode is a well-established fingerprint of the charge-density-wave order, the nature of the sharp sub-gap mode emerging below the superconducting temperature is still under debate. In this work we use external pressure as a knob to unveil the delicate interplay between the two orders, and consequently the nature of the superconducting mode. Thanks to an advanced extreme-conditions Raman technique we are able to follow the pressure evolution and the simultaneous collapse of the two intertwined charge-density-wave and superconducting modes. The comparison with microscopic calculations in a model system supports the Higgs-type nature of the superconducting mode and suggests that charge-density-wave and superconductivity in 2H-NbSe$_2$ involve mutual electronic degrees of freedom. These findings fill the knowledge gap on the electronic mechanisms at play in transition metal dichalcogenides, a crucial step to fully exploit their properties in few-layers systems optimized for devices applications.


\end{abstract}

\maketitle


\section{\label{introduction}Introduction}

The symmetry breaking across an electronic phase transition always occurs along with the emergence of new collective excitations. The charge-density-wave (CDW) electronic instability is accompanied by the softening of a phonon coupled to the electronic density at ${\bf Q}_{CDW}$ and dressed by the amplitude fluctuations of the CDW order parameters which develops below T$_{CDW}$\cite{rice_ssc74,gruner_review}. This new mode, also called amplitudon, is Raman active\cite{klein_raman82,cea_cdw_prb14} and has been detected in several CDW dichalcogenides, including 2H-NbSe$_2$\cite{tsang_prl76,measson_prb14,mak_natnano15}.

In the superconducting (SC) state, two additional collective excitations of the superconducting order parameter 
are expected: a massless Nambu-Goldstone phase mode, which is pushed to the plasmon frequency in a charged superconductor, and a massive amplitude mode, also named Higgs mode for the analogy with the Higgs boson in high-energy physics\cite{nagaosa}.
In principle, the Higgs mode remains `dark' to spectroscopy probes, since it weakly couples  to the electromagnetic field\cite{cea_thg_prb16} and is overdamped\cite{volkov73,kulik81,varma_prb82,cea_cdw_prb14,cea_prl15}. Indeed its energy coincides with the threshold $2\Delta_{SC}$ of the quasiparticle continuum, $\Delta_{SC}$ being the superconducting gap\cite{volkov73,kulik81,varma_prb82,varma_higgs_2002,cea_cdw_prb14}. Even though some recent reports investigated the possibility to detect it via optical spectroscopy in strongly-disordered superconductors\cite{frydman_natphys15} or intense THz field\cite{shimano_prl13,shimano_science14}, its presence and observability are still under strong debate\cite{pekker_amplitudehiggs_2015,cea_prl15,cea_thg_prb16,armitage_prb16,tsuji_prb16}.

On the other hand, when superconductivity coexists with a CDW order the Higgs mode has the unique opportunity to become visible via its coupling to the soft CDW phonon mode. 2H-NbSe$_2$ is one of the few systems where the two orders coexist, with a CDW and superconducting instabilities at $T_{CDW}=33$~K and $T_c=7$~K, respectively. This mechanism has been proposed long ago\cite{varma_prb82, littlewood_gauge-invariant_1981} to explain the dramatic changes of the Raman spectrum of 2H-NbSe$_2$ below $T_c$, where a sharp peak develops below $2\Delta_{SC}$ by stealing spectral weight from the soft phonon peak \cite{klein_prb81,measson_prb14}. Nonetheless, other mechanisms could give rise to sharp superconducting resonances, as observed in other multiband superconductors\cite{blumberg_prl07,hackl_prx14}, making the assignment of the superconducting peak to the Higgs mode problematic.\\

\begin{figure*}[ht]
\centering
\includegraphics[width=\linewidth]{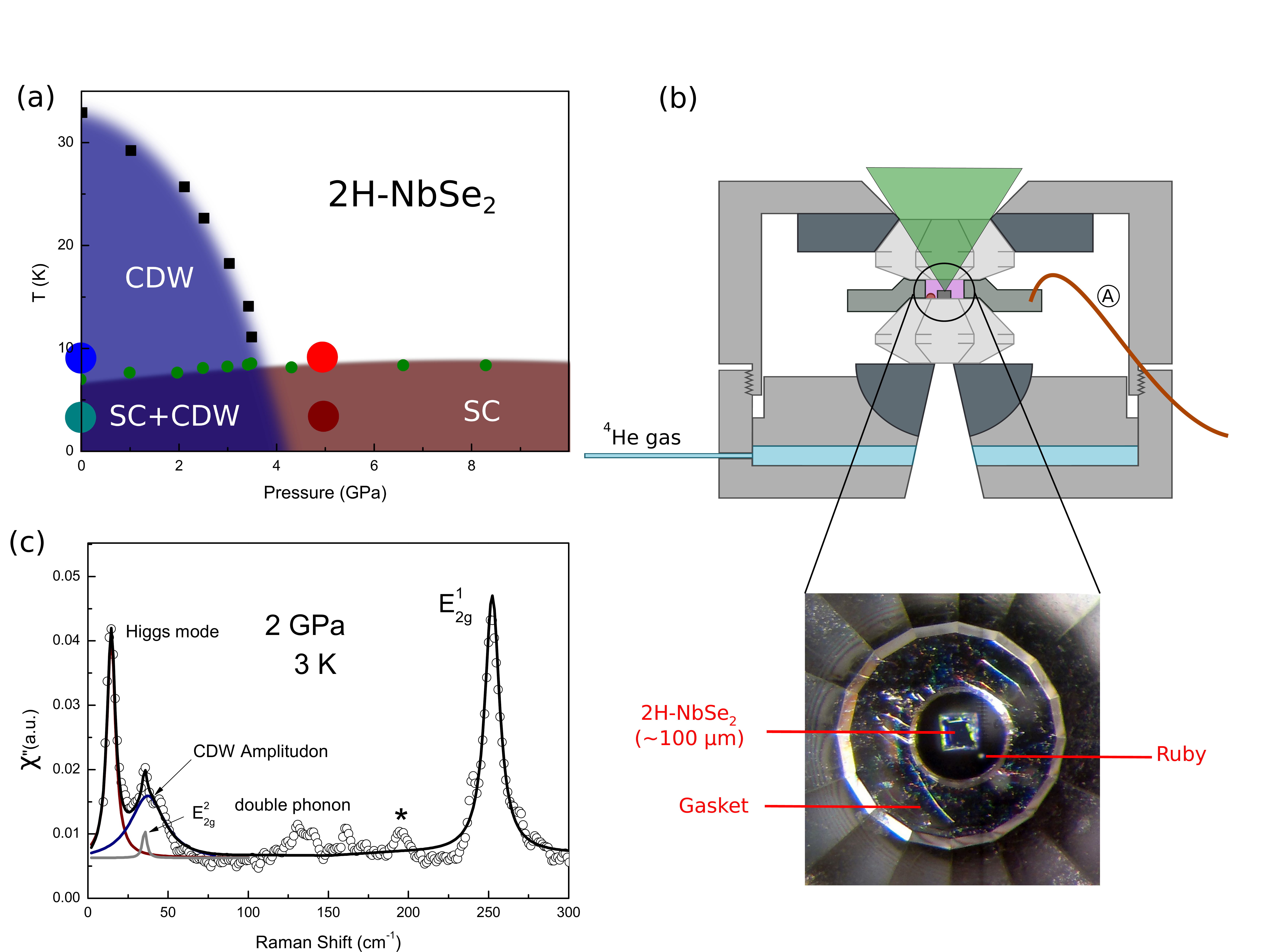}
\caption{Raman scattering under hydrostatic pressure and at low temperature of 2H-NbSe$_2$. (a) (P,T) Phase diagram of 2H-NbSe$_2$ drawn from resistivity measurements\cite{suderow_pressure_2005,jerome_layer_1976}. The incommensurate CDW collapses at a critical pressure of $\sim$~4~GPa, a pure SC state persists up to at least 10~GPa. Large colored circles mark the (P,T) positions of the experimental spectra reported in Fig.~\ref{fig:2}(a). (b) Membrane Diamond anvil cell designed for a large numerical aperture collection (green cone), low Raman signal from the environment and access to low temperature ($\sim$~3~K). The 350 $\mu$m diameter pressure chamber containing the freshly cleaved 2H-NbSe$_2$ sample and rubies as pressure gauge is depicted below. (A) Thermal link between the metallic gasket and the cold finger of the cryostat made of high conductivity Cooper wires. (c) Raman spectra of 2H-NbSe$_2$ in the E$_{2g}$ symmetry at 2~GPa and 3~K in the coexisting region of charge-density-wave and superconducting orders. The $\star$ designs a CDW-phonon mode. Consistently with the theory, the two collective excitations at 14~cm$^{-1}$ and 40~cm$^{-1}$ at 2~GPa are assigned to the superconducting Higgs mode and the charge-density-wave amplitudon mode.}
\label{fig:1}
\end{figure*}

Tuning the delicate interplay between the CDW and superconductivity is achievable by application of high pressure or by lowering the dimensionality of the system. Pressure experiments in 2H-NbSe$_2$ showed that above a critical pressure of 4~GPa the CDW order disappears while superconductivity remains almost unaffected\cite{jerome_ssc76,suderow_pressure_2005,feng_pnas12,rodiere_prb15} (Fig.~\ref{fig:1}(a)). This behavior is in striking contrast to what is found by reducing the sample dimensionality since, there, CDW order is significantly reinforced while superconducting transition temperature is halved for monoloayer system\cite{mak_natnano15,staley_2009,Ugeda_NP_2015}. All these observations triggered intense theoretical efforts\cite{mazin_prb06,mauri_prb09,zhu_pnas15,flicker_natcomm15,Weber_prb16} to explain the origin of the CDW and superconducting transitions in bulk and few-layers 2H-NbSe$_2$, by accounting for the different role of the electron-phonon coupling and Fermi-surface nesting. So far, pressure effects have been addressed via signatures of the lattice and its dynamics \cite{feng_pnas12,rodiere_prb15}. Raman probe under pressure has the advantage to assess directly the evolution of the {\em electronic} degrees of freedom with the pressure-driven CDW softening, without the additional complication of the interaction with the substrate, relevant in devices based on few-layers 2H-NbSe$_2$.

Here we use an advanced low-temperature and high-pressure technique to probe the Raman signatures of the CDW and superconducting excitations across the phase diagram of 2H-NbSe$_2$. Besides, we compute the evolution of the Raman response within a microscopic model for the coexisting CDW and superconducting order. Our findings point to the assignment of the SC peak to the Higgs fluctuations as the most likely interpretation. This result not only provides a perspective for the microscopic mechanisms at play in the coexisting states of 2H-NbSe$_2$, but it also shows that the Higgs-mode radiance is a direct fingerprint of charge ordering.

\section{Experimental details}

Single crystals of 2H-NbSe$_2$ were synthesized at 750$^\circ$C using the iodine-vapour transport method as described elsewhere\cite{fisher_stoichiometry_1980}. The crystallographic quality of several crystals was checked by X-ray diffraction. All crystals revealed a hexagonal cell with parameters a=b=3.44$\angstrom$ and c=12.54$\angstrom$ in agreement with the 2H polytype.

We have adapted an original optical experimental setup (Fig.~\ref{fig:1}(b)) to probe low-energy Raman excitations under extreme conditions of pressure and low temperature. This report constitutes the first experiment of this type, successfully reaching low-energy down to 7~cm$^{-1}$ (0.85~meV), down to $\sim$~3~K and up to 5~GPa. We have thus been able to track simultaneously the superconducting and the CDW modes under a broad range of hydrostatic pressure.
Raman scattering measurements have been performed on freshly cleaved 2H-NbSe$_2$ crystal with an incident angle of $\sim$~30$^\circ$ with respect to the sample surface normal and in a membrane diamond anvils cell designed for a large numerical aperture together with a low Raman signal from the environment of the sample as described in \cite{buhot_driving_2015}. The pressure cell was cooled down in a closed-cycle $^4$He cryostat with a base temperature of 3~K. As sketched in Fig.~\ref{fig:1}(b), a OFHC Cooper braid between the metallic gasket and the cold finger is used as a cryogenic leakage. This, together with a low laser incident power (typically 0.1~mW) and control of the size of the incident laser spot (about 20~$\mu$m diameter), allowed to reach low temperature, estimated between 3 and 3.5~K, low-enough to measure a Raman signature of the superconducting state of 2H-NbSe$_{2}$ ($T_c$ = 7.2~K).

 We have used a triple-grating spectrometer Jobin Yvon T64000 equipped with a liquid-nitrogen-cooled CCD detector and the 532~nm excitation line from a solid-state laser.  
The polarization of the incoming and outgoing light are in the (ab) plane of the sample. In this configuration, in parallel and crossed polarizations we select the A$_{1g}$+E$_{2g}$ and the E$_{2g}$ symmetries, respectively. Birefringence of the diamonds under pressure mixes the effective polarization of lights. Extraction of pure E$_{2g}$ and $A_{1g}$ symmetries was done by scaling the E$_{2g}$ phonon mode at about 250~cm$^{-1}$. The fluorescence of ruby has been used as a pressure gauge. The pressure transmitting medium is $^4$He. It does not show any particular Raman features at low energy (down to 7~cm$^{-1}$), down to 3~K and up to 10~GPa.

\section{Collapse of the collective modes}

A typical spectrum of 2H-NbSe$_2$ under high pressure (2~GPa) and at low temperature in the coexisting region of charge-density-wave (CDW) and superconducting (SC) states is displayed in Fig~\ref{fig:1}(c) in a large Raman shift range. In the E$_{2g}$ symmetry, beyond the single phonons E$^2_{2g}$ and E$^1_{2g}$ (at 26 and 252~cm$^{-1}$, respectively), a second-order phonon peak ($\sim$~130 cm$^{-1}$) and a CDW-phonon ($\sim$~200 cm$^{-1}$, marked with \textbf{$\star$}) are observed. The last one, also observed in the A$_{1g}$ symmetry, is a signature of the CDW ordering, most probably a phonon mode folded to the zone center due to the CDW ordering. It remains a hard mode up to T$_{CDW}$ while the low energy collective modes at $\sim$~15~cm$^{-1}$ and $\sim$~40~cm$^{-1}$ soften upon approaching the ordering temperatures T$_{c}$ and T$_{CDW}$, respectively. Both these SC and CDW collective modes are visible in the fully symmetrical A$_{1g}$ (CDW soft mode at 39~cm$^{-1}$ and SC mode at 19~cm$^{-1}$ at ambient pressure) and the in-plane symmetry breaking E$_{2g}$ (CDW soft mode at 43.5~cm$^{-1}$ and SC mode at 14~cm$^{-1}$ at ambient pressure) channels. The CDW soft mode has been already identified in previous literature\cite{tsang_prl76,measson_prb14} with the so called amplitudon, i.e. the instability phonon dressed by amplitude fluctuations of the CDW.  As we will discuss below, we assign the SC mode in the coexisting region of CDW and SC states to a signature of the amplitude fluctuations of the SC order parameter, so we will denote it as Higgs mode in what follows. As already discussed ~\cite{measson_prb14}, a partial spectral weight transfer from the CDW soft mode to the SC Higgs mode exists with decreasing temperature.  While the Higgs mode rises, the CDW soft mode looses spectral weight, as observed in the two different symmetries E$_{2g}$ and A$_{1g}$ (see also Fig.~\ref{fig:5}).\\
\begin{figure*}[ht]
\centering
\includegraphics[width=\linewidth]{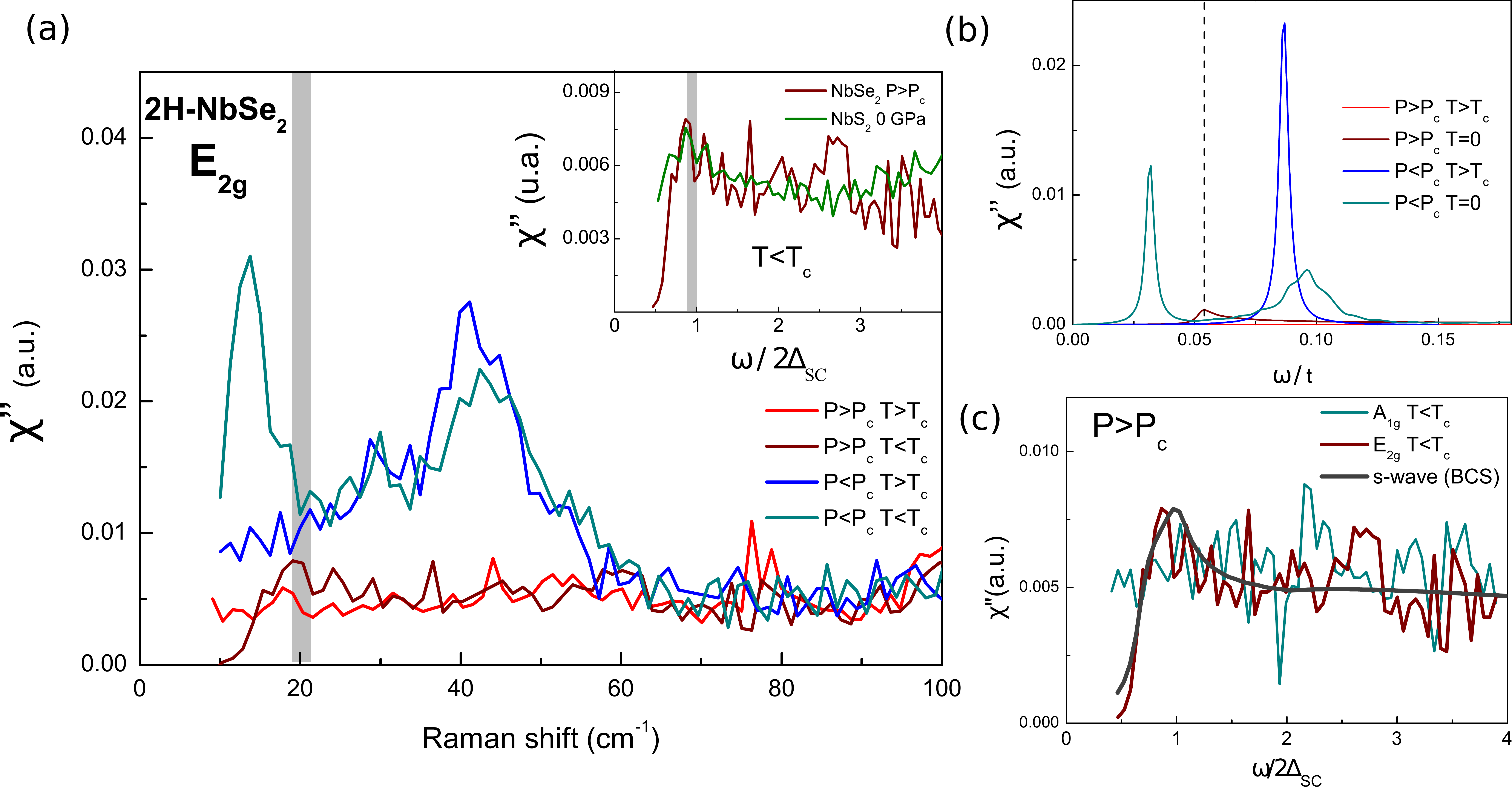}
\caption{Collapse of the superconducting Higgs mode in the pure superconducting state of 2H-NbSe$_2$, measurements and theoretical predictions. (a) Raman spectra in the E$_{2g}$ symmetry measured at various (P,T) positions as identified Fig.~\ref{fig:1}(a): in the coexisting SC+CDW (green) and pure CDW (blue) states at ambient pressure and in the pure SC (brown) and paramagnetic (red) state at high pressure. Both the CDW amplitudon and the SC Higgs modes disappear at high pressure. A small Cooper-pairs breaking peak remains at $2\Delta_{SC}$. $2\Delta_{SC}$ is marked by the grey band ranging from $2\Delta_{SC}$ measured by STM \cite{guillamon_intrinsic_2008} at ambient pressure to the value we extrapolate at high pressure accordingly to the increase of T$_c$ with pressure \cite{suderow_pressure_2005}. Inset: Raman spectra in the pure SC state of 2H-NbSe$_2$ (above 4~GPa) and non-CDW NbS$_2$ (0~GPa) versus the Raman shift normalized to $2\Delta_{SC}$ \cite{guillamon_intrinsic_2008,guillamon_superconducting_2008}.
(b) Theoretical Raman responses calculated in a microscopic model (see text and Appendix \ref{model}) in the four phases (SC+CDW, CDW, SC,PM) for comparison with the experimental spectra in (a). $t$ is the hopping term. The parameters are: $\Delta_{SC}/t$=0.025, $g_{CDW}/t$=0.14 and 0.12 in the SC+CDW and SC phases, respectively. The spectra are well-reproduced in all phases. (c) Raman response of 2H-NbSe$_2$ in the pure superconducting phase above P$_c$ in the E$_{2g}$ and A$_{1g}$ symmetries. The black line is the theoretical Raman response of a Cooper-pairs breaking peak in a two-gaps (or anisotropic gap) s-wave superconductor in the BCS regime with an additional electronic background $\beta$. The form of $\beta$ is :  $\beta(\omega)=a\omega/\sqrt{b+c\omega^2}$. It barely affects the shape of the Cooper-pairs breaking peak.}
\label{fig:2}
\end{figure*}

By applying high pressure above 4~GPa (Fig.~\ref{fig:1}(a)) a pure superconducting state is reached and, as presented in Fig.~\ref{fig:2}(a), both the CDW amplitudon and the Higgs mode disappear. On the other hand, in the E$_{2g}$ symmetry a weak SC signature persists, with marked differences with respect to the sharp SC Higgs mode seen at $P<P_c$=4~GPa. Indeed, its intensity is a factor of $\sim 8$ smaller and its energy suddenly hardens. As shown in the inset of Fig.~\ref{fig:2}(a), the superconducting Raman response of the compound 2H-NbS$_2$, which lacks the CDW state at ambient pressure \cite{measson_prb14}, matches perfectly that of 2H-NbSe$_2$ above the critical pressure P$_c$, in the pure superconducting state, as long as the Raman shift is scaled by $2\Delta_{SC}$ ($\Delta_{SC}$ is calculated from the value of the superconducting gap measured by STM~\cite{guillamon_scanning_2008} and its pressure dependence is scaled as $T_c$(P)\cite{suderow_pressure_2005}). Both superconducting thresholds are positioned at $2\Delta_{SC}$ as expected for a simple Cooper-pairs breaking peak.
As pointed out by many recent measurements\cite{noat_quasiparticle_2015,rahn_gaps_2012,fletcher_penetration_2007,jing_quasiparticle_2008,rodrigo_stm_2004,borisenko_two_2009}, 2H-NbSe$_2$ is an s-wave superconductor with either an anisotropic gap or multiple gaps. This property affects the shape of the Raman Cooper-pairs breaking peak.  As shown in Fig.~\ref{fig:2}(c), the Raman spectrum can be properly reproduced by defining an anisotropic gap which varies from a minimum value of $\Delta_{SC}^s$=0.92~meV to a maximum value of $\Delta_{SC}^L$=1.38~meV \cite{guillamon_intrinsic_2008}. We cannot distinguish here between the presence of multiple gaps or a single anisotropic gap and we do not exclude $k_z$ dependency in the real compound 2H-NbSe$_2$ \cite{Weber_prb16,noat_quasiparticle_2015}. Mainly our fit provides evidence for the nature of the SC peak above P$_c$, i.e. a Cooper-pairs breaking peak, with insight into the energy scale of the superconducting gap.
Consistently with this assignment,  in the A$_{1g}$ symmetry there is no signature of the pure superconducting state reached in 2H-NbSe$_2$ above the critical pressure, due to Coulomb screening effect \cite{devereaux_electronic_1995,devereaux_inelastic_2007,cea_raman_prb16} (see Fig.~\ref{fig:2}(c)).

The disappearance of the sharp SC mode below $2\Delta_{SC}$ in the pure superconducting phase demonstrates unambiguously its intimate link with the coexisting charge-density-wave order. These findings, consistently with the theory discussed below, support the Higgs type assignment of the sharp SC mode below P$_c$.

\section{Comparison with a microscopic model}

In 2H-NbSe$_2$ the phonon coupled to the CDW belongs to an acoustic branch, so the single-phonon mode is not visible as a finite-energy peak in ${\bf q}$$\sim$~0 Raman spectroscopy above $T_{CDW}$. Below $T_{CDW}$  the intermediate electron-hole excitations which couple directly to light allow to make the phonon mode at ${\bf Q}_{CDW}$ Raman visible at ${\bf q}=0$. This gives rise to the soft phonon modes at $\sim~40~$cm$^{-1}$. In a general approach, the Raman response below $T_{CDW}$ can be schematically written as
\begin{equation}
\label{chi}
\chi"(\omega)=Z_{eff}(T,\Delta_{CDW}) \frac{\Gamma_{ph}}{(\omega^2-\Omega_0^2(T))^2+\Gamma_{ph}^2}
\end{equation}
where the soft mode frequency $\Omega_0(T)$ and damping $\Gamma_{ph}$ are both determined by the CDW amplitude fluctuations and the prefactor $Z_{eff}\sim \Delta_{CDW}^2$ grows proportionally to the CDW order parameter\cite{cea_cdw_prb14}. The frequency $\Omega_0(T)$ also scales approximately with the CDW gap, so it goes to zero at $T_{CDW}$, even though the Raman peak disappears already at $T\simeq 0.9 T_{CDW}$ due to the strong suppression of $Z_{eff}$.
While the assignment of the soft CDW peak to the amplitudon is well established in the literature\cite{tsang_prl76,varma_prb82,klein_raman82,measson_prb14,mak_natnano15,Weber_extended_2011} the interpretation of the additional peak emerging upon entering the SC state has been somehow controversial.  The first suggestions\cite{balseiro,varma_prb82} assumed that the amplitudon can be treated as an ordinary ${\bf q}=0$ Raman-active soft phonon, and considered how the proximity of $\Omega_0$ to the scale $2\Delta_{SC}$ can modify the  phonon spectral function itself. Balseiro and Falikov\cite{balseiro} proposed that the SC peak originates from an ordinary self-energy correction of the phonon due to the coupling to electronic excitations, whose quasiparticle spectrum changes after the gap opening\cite{Zeyher_Superconductivity_1990}. This mechanism is analogous to the one proposed to interpret the changes in the lineshape of finite-momentum strongly-damped phonons  measured by neutron scattering in systems like YNi$_2$B$_2$C or LuNi$_2$B$_2$C\cite{allen_prb97,kawano_prl96,weber_prl08}.

 However, as correctly pointed out by Littlewood and Varma later on\cite{varma_prb82}, a ${\bf q}=0$  symmetric (A$_{1g}$) phonon couples also to the long-range Coulomb interactions\cite{Zeyher_Superconductivity_1990}, which renormalize to zero the self-energy phononic corrections in the particle-hole channel. In contrast, if the soft phonon is coupled to the Higgs fluctuations, the self-energy corrections due to SC amplitude fluctuations are not affected by the Coulomb screening. Then this mechanism can lead to sharp sub-gap peaks,   even in the A$_{1g}$ symmetry. In the case of this last scenario, the SC signature is not the pure Higgs mode. Rather we observe its manifestation on the CDW amplitudon, which is split thanks to the interaction with the Higgs fluctuations.
 
The milestone idea by Littlewood and Varma has been put later on firmer grounds\cite{browne_prb83,cea_cdw_prb14}. Browne and Levin\cite{browne_prb83} explained that the coupling between the soft CDW phonon and the Higgs fluctuations originates microscopically from the intertwined amplitude fluctuations of the two CDW and SC order parameters. More recently, Cea and Benfatto\cite{cea_cdw_prb14} computed explicitly the Raman response, evaluating the intermediate electron-hole processes which make the CDW phonon Raman visible, i.e. the effective charge $Z_{eff}$ in Eq.\ (\ref{chi}) above. The microscopic identification of the coupling between the amplitudon and the Higgs implies  that the CDW and SC order parameters should overlap at least in part of the Fermi surface, so that their amplitude fluctuations talk to each other via a modification of the same electronic density-of-states.  As it has already been shown in Ref.\ \cite{cea_cdw_prb14}, the calculation of the Raman response  within a microscopic model system for the coexisting state is able to reproduce successfully the main feature of the experiments at ambient pressure. In addition such a microscopic approach clarifies that accounting only for the change in the particle-hole spectrum of excitations due to the superconducting gap opening is not enough to reproduce the strong sub-gap peak (see Appendix \ref{model} for further details). Indeed, in contrast to the usual case of a metallic-to-superconductor transition considered e.g. in Ref. \cite{balseiro,allen_prb97,Zeyher_Superconductivity_1990}, here the quasiparticle spectrum above $T_c$ is already gapped by the CDW gap, being then weakly affected by the opening of the superconducting one. Thus the changes in the phonon lineshape when going from the CDW to the superconducting state cannot be simply ascribed to a redistribution of the charge excitations across $2\Delta_{SC}$, as described in the previous work\cite{balseiro,allen_prb97,Zeyher_Superconductivity_1990} focusing on standard phonon. Besides, even if considered as an ordinary phonon mode, the soft mode in 2H-NbSe$_2$ is at $\sim 2\cdot2\Delta_{SC}$ and its tail does not overlap with $2\Delta_{SC}$. So the mechanism of spectral-weight redistribution around $2\Delta_{SC}$ \cite{balseiro,allen_prb97,Zeyher_Superconductivity_1990} fails to reproduce the intense sub-gap peak, even in the E$_{2g}$ symmetry.

 Here, following the  approach of Ref.\ \cite{cea_cdw_prb14}, we model the pressure effects by a continuous suppression of the couplings in the CDW and superconducting channels, in order to reproduce the suppression of the CDW gap while keeping $\Delta_{SC}$ almost constant (see Appendix \ref{model} for further details). As a control parameter playing the role of the pressure we then use the relative change $\alpha=2(g_{CDW}^0-g_{CDW})/g^0_{CDW}$ of the CDW coupling $g_{CDW}$.  The Raman intensities are then computed in the various phases (pure SC, CDW+SC, CDW such as measured) (see Fig.~\ref{fig:2}(b)). The spectra have the same absolute units, so the scaling of the intensities is respected. The theoretical Raman response is consistent with our measurements: the Higgs mode manifests as a secondary peak of the CDW soft phonon, which is the mode Raman visible. Thus it appears as a sub-gap intense peak only when it coexists with a CDW state. When the CDW disappears the Raman response in the pure superconducting state displays only a broad and weak signature at 2$\Delta_{SC}$. Besides, the hardening and damping of the amplitudon mode upon entering the SC state is a direct consequence of its coupling to the collective electronic excitations, whose density-of-states gets redistributed from below to above $2\Delta_{SC}$ in the superconducting state. It is actually observed experimentally at all pressures below 4~GPa (see Fig.~\ref{fig:4}(a,b)).
\begin{figure*}[ht]
\centering
\includegraphics[width=15cm]{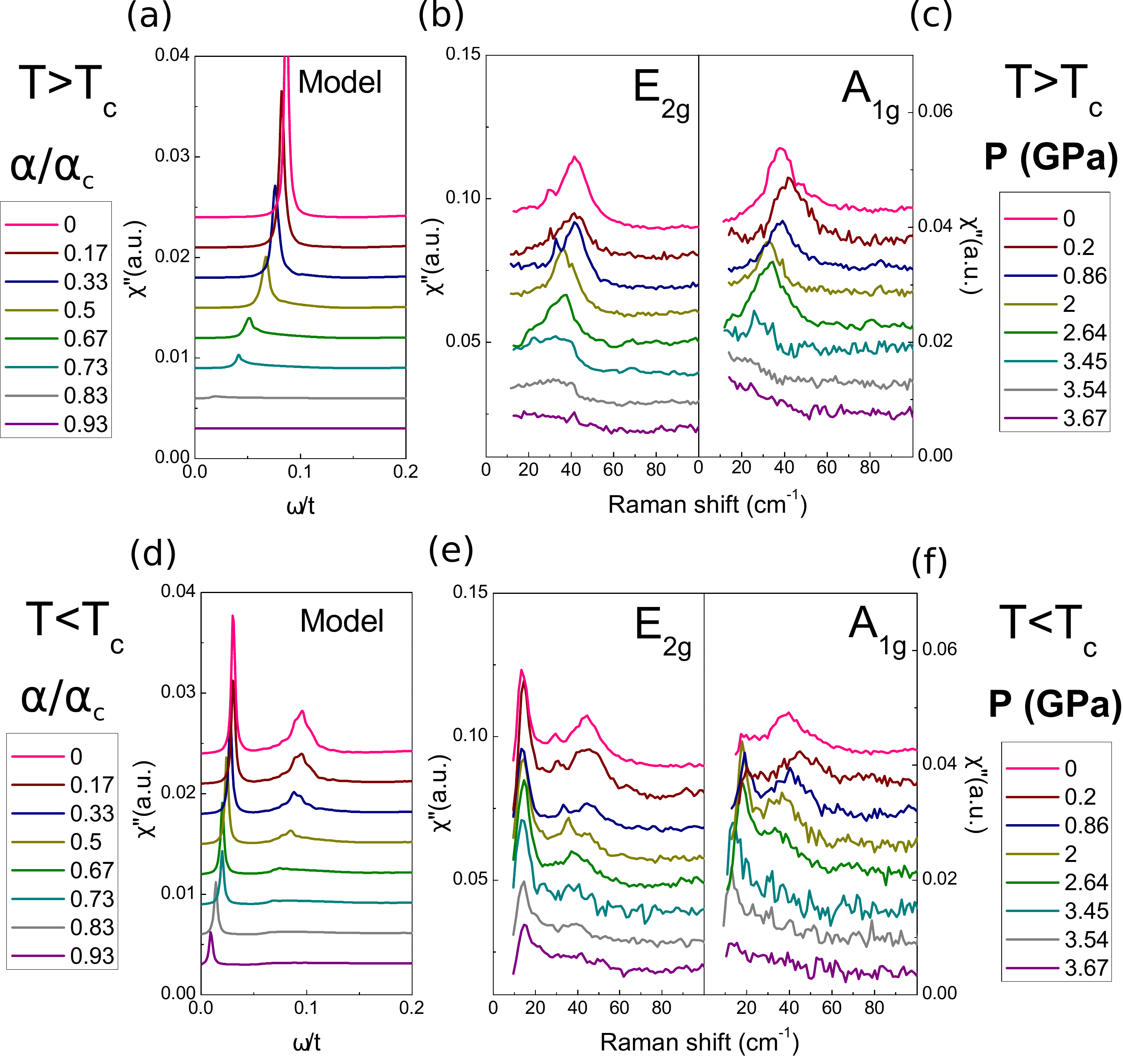}
\caption{Pressure dependence of the Raman spectra of 2H-NbSe$_2$ in the coexisting superconducting and charge-density-wave phases, experiments and theory. Raman response at  8~K (b,c) in the CDW state and at 3~K (e,f) in the CDW+SC state for various pressures up to the critical pressure. The spectra are normalized to the E$_{2g}$ phonon and consecutively shifted up. (a,d) Theoretical Raman response computed microscopically, with frequency given in units of the hopping parameter $t$, which sets the energy scale. The  pressure dependence is simulated by suppressing the CDW coupling $g_{CDW}$ with respect to its value $g_{CDW}^0$ at ambient pressure, with $\alpha=2(g_{CDW}^0-g_{CDW})/g^0_{CDW}$. The experiments at a given $P/P_c$ are compared to calculations at $\alpha/\alpha_c$, where $\alpha_c=0.30$ is the critical coupling at which CDW order disappears.} 
\label{fig:3}
\end{figure*}

\section{Tuning the interplay between CDW and SC with pressure}
To further unveil the interplay between CDW and SC in 2H-NbSe$_2$ we have finely tuned the pressure in the coexisting CDW+SC state, both experimentally and theoretically. Figure~\ref{fig:3}(b,c) and \ref{fig:3}(e,f) reports the experimental results in the E$_{2g}$ and A$_{1g}$ symmetries above and below T$_{c}$, respectively, from ambient pressure up to 3.67~GPa, corresponding to $P/P_c=0.92$. The intensities are normalized on the high energy E$_{2g}$ phonon mode. As one can see in the upper panels, the CDW amplitudons gradually soften, enlarge and loose intensity with increasing pressure. At $P=3.54$~GPa ($P/P_c=0.89$), the amplitudon is barely visible above $T_c$, even though the critical pressure has not been reached yet. In contrast, as shown Fig.~\ref{fig:3}(e,f), the SC Higgs peaks are visible up to $P_c$, leading to the remarkable effect that the radiance of the SC Higgs signature guarantees that a residual CDW order is present. At the same time, as the system is cooled below T$_c$, the amplitudon shifts to higher energy and gets enlarged, demonstrating a clear coupling between the two peaks.
All these features are well reproduced by our calculations shown in Fig.~\ref{fig:3}(a,d). As we mentioned before, the CDW instability is progressively suppressed as the relative CDW coupling $\alpha$ increases, up to the critical value $\alpha_c=0.3$ where it disappears. We then compare the experimental results for $P/P_c$ to our calculations at the corresponding $\alpha/\alpha_c$.  In the model, the softening of the energy of the amplitudon  for increasing $\alpha$ is due to the suppression of the CDW gap, since the CDW amplitude fluctuations are peaked at $2\Delta_{CDW}$. Simultaneously the Raman intensity $Z_{eff}\sim \Delta^2_{CDW}$ is rapidly suppressed, making the Raman signature of the CDW amplitudon above $T_{c}$ barely visible already at $\alpha/\alpha_c=0.8$, in agreement with the experiments. On the other hand, since the Higgs mode is much sharper than the amplitudon at any pressure, even in this regime near $\alpha_c$, it is clearly visible, giving a clear fingerprint of the existence of a CDW order.

\begin{figure*}[ht]
\centering
\includegraphics[width=\linewidth]{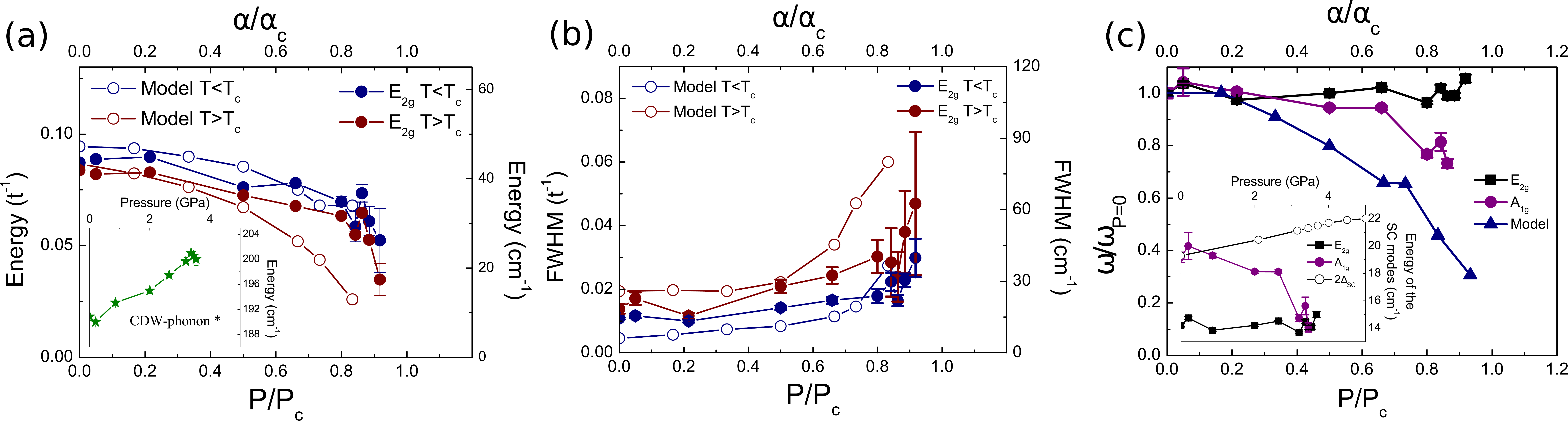}
\caption{(a) Pressure evolution of the energy of the amplitudon in the E$_{2g}$ symmetry above and below T$_{c}$ compared to the evolution of the amplitudon in the microscopic model. Inset: Evolution of the energy of the folded CDW phonon,  denoted with * (190~cm$^{-1}$) in the spectra of Fig~\ref{fig:1}(b), measured at 8~K. (b) Pressure evolution of the width of the amplitudon in the E$_{2g}$ symmetry above and below T$_{c}$ compared to the evolution of the amplitudon in the microscopic model. (c) Pressure dependence of the energy of the Higgs mode normalized to its  value at zero pressure in the two A$_{1g}$ and E$_{2g}$ channels, and in the microscopic model. From 0 to P$_c$=4~GPa, the A$_{1g}$ mode softens, qualitatively following the behavior of the Higgs mode in the microscopic model, whereas the E$_{2g}$ one is constant, showing strong symmetry-dependent behavior. Inset: Pressure dependence of the Higgs mode in both symmetries compared to the pair-breaking threshold $2\Delta_{SC}$, with $\Delta_{SC}$ extrapolated from STM measurements\cite{guillamon_intrinsic_2008} at ambient pressure scaled with the pressure evolution of T$_c$\cite{suderow_pressure_2005}.}
\label{fig:4}
\end{figure*}

The detailed pressure dependence of the energy and width of both the amplitudon and the Higgs mode is reported in Fig.~\ref{fig:4}. The CDW amplitudons gradually soften with increasing pressure and harden upon entering the SC state at all pressures. This tendency is well reproduced by our microscopic model (see Fig.~\ref{fig:4}(a)). For the sake of completeness we also show in the inset of Fig.~\ref{fig:4}(a) the pressure dependence of the folded CDW-phonon mode at $\sim~190~cm^{-1}$ (marked with \textbf{$\star$} in Fig.~\ref{fig:1}(c)). As one can see, in contrast to the amplitudons, its energy hardens linearly with pressure in the same way the regular A$_{1g}$ and E$_{2g}$ phonons do. More precisely, the rate of increase of the energy of the A$_{1g}$, E$^1_{2g}$ and the folded CDW-phonon mode is similar at about 1$\%$ per GPa.
 In panel~\ref{fig:4}(b) we compare the pressure evolution of the width of the amplitudon above and below $T_c$ with the theoretical calculations. The experimental trends are very well captured by the model, with an overall broadening of the amplitudon upon entering the SC state at a given pressure or as the pressure increases. The larger variations in temperature found theoretically can be ascribed to the presence of a sharper phonon peak in the calculations above $T_c$ (see also Fig.~\ref{fig:3}(a)). Since the broadening of the phonon peak is provided by residual quasiparticle scattering events from ungapped regions of the Fermi surface, it crucially depends on details of the band structure in the CDW state.

In panel~\ref{fig:4}(c) we summarize the evolution of the SC Higgs peaks energy in the two channels. As highlighted in the inset, the Higgs mode always lies below $2\Delta_{SC}$, where the usual Cooper-pairs breaking peak would be instead expected.
In addition, the A$_{1g}$ peak softens by 30~$\%$ and the energy of the E$_{2g}$ peak stays constant whereas the critical temperature $T_{c}$ rises with pressure. The absence of scaling between the SC peaks energy and T$_{c}$ in the SC+CDW coexisting phase indicates that, in both symmetries, the SC mode below P$_{c}$ is not simple a Cooper-pairs breaking peak. Moreover, this is an additional evidence that the SC mode, even in the E$_{2g}$ symmetry where Coulomb screening is not effective, is not a SC peak originates from an ordinary\cite{balseiro,allen_prb97,Zeyher_Superconductivity_1990} self-energy correction of the phonon,  since this mechanism would predict that the peak position follows the pressure evolution of $2\Delta_{SC}$.

While in the A$_{1g}$ channel the Higgs mode has the same pressure trend than our calculations, in the E$_{2g}$ symmetry the Higgs mode energy shows a relatively different behavior. To understand such a discrepancy one should notice that in our simplified model the CDW instability occurs at a single ${\bf Q}_{CDW}$ vector equivalent to half of the reciprocal-lattice wavevector. In this situation, the CDW phonon is only visible in the fully symmetrical A$_{1g}$ channel. On the other hand, in 2H-NbSe$_2$ the CDW instability can occur at three equivalent ${\bf Q}^i_{CDW}$ wavevectors connected by a $2\pi/3$ rotation. This guarantees that the amplitudon has a finite projection in both A$_{1g}$ and E$_{2g}$ symmetries\cite{klein_raman82}. When the SC state forms the Higgs fluctuations renormalize the frequencies of the phonon modes corresponding to lattice displacements at the three ${\bf Q}^i_{CDW}$ wavevector. If the electron-phonon coupling  has a non-trivial momentum dependence, as it has been emphasized recently\cite{mauri_prb09,flicker_natcomm15}, one cannot exclude a splitting of the Higgs signatures, which reflects in a different trend of the subgap peak observed in the A$_{1g}$ and E$_{2g}$ channels under pressure. A full understanding of this issue requires a calculation within a microscopic model for 2H-NbSe$_2$, which is beyond the scope of the present paper.

\begin{figure*}[ht]
\centering
\includegraphics[width=\linewidth]{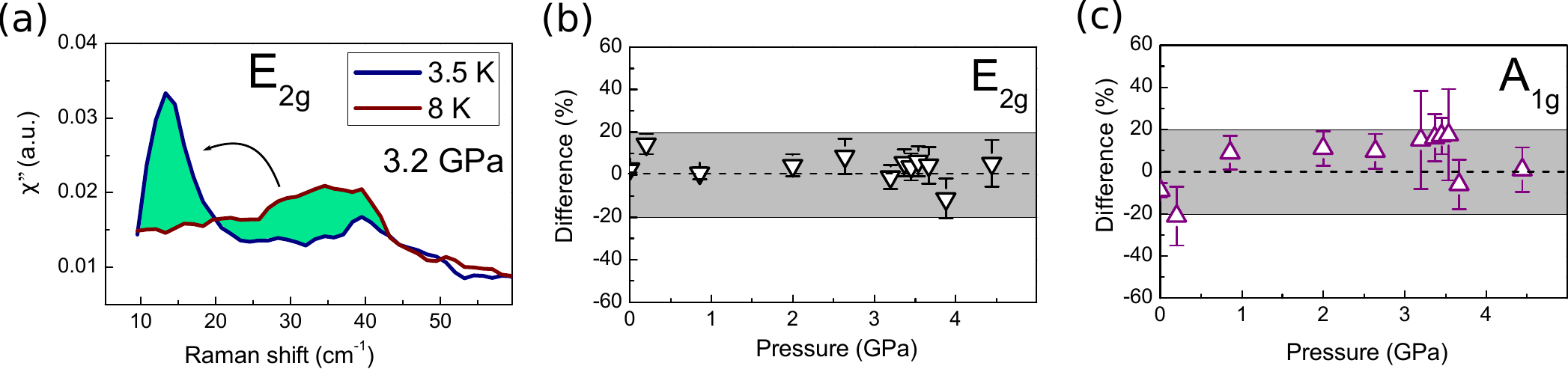}
\caption{Spectral weight transfer between the superconducting mode and the charge-density-wave mode. (a) Raman spectra of 2H-NbSe$_2$ at 3.2~GPa above and below T$_{c}$ in the E$_{2g}$ symmetry. The spectral weight transfer is depicted in green. (b,c) Difference (in $\%$) of the total spectral weight of both modes between 3~K and 8~K (below/above T$_{c}$) as a function of pressure in E$_{2g}$ (b) and A$_{1g}$ (c) symmetries. At every pressure and in both symmetries, the total spectral weight of the Higgs mode and the amplitudon is relatively conserved ($\sim20\%$) within the large error bars.}
\label{fig:5}
\end{figure*}

Finally, in Fig.~\ref{fig:5} we show in details the transfer of spectral weight between the amplitudon and the Higgs mode below $P_c$. Even though in Raman spectroscopy the total spectral weigh is not constrained by a sum rule, as it happens for optical spectroscopy, previous work at ambient pressure has shown \cite{measson_prb14} that upon entering the superconducting state the rise of the Higgs mode happens at the expense of the CDW amplitudon. This finding is also observed for increasing pressure up to the collapse of both modes, as shown in Fig.~\ref{fig:5}(b) and (c) for the E$_{2g}$ and A$_{1g}$ symmetries, respectively. The total spectral weight of the two modes is approximately conserved (at $\pm$~20$\%$) when the system is cooled below $T_c$, even though error bars here are larger as compared to experiments at ambient pressure \cite{measson_prb14}. This approximate conservation of spectral weight further demonstrates the existence of a direct coupling between the SC Higgs mode and the CDW amplitudon at every pressure below $P_c$.

\section{\label{discussion}Discussion}

The comparison between the theoretical calculations and the experiments points to the assignment of the SC mode in the CDW and SC coexisting region to a signature of the Higgs mode carried out by the CDW amplitudon as the most likely. In particular, the sudden disappearance of the SC signature as $\Delta_{CDW}=0$ agrees with the general prediction of Eq.\ (\ref{chi}) that its Raman visibility is only guaranteed by the presence of a soft, Raman-active CDW amplitudon, allowing for $Z_{eff}\neq 0$. Indeed, other possible interpretations based on the multiband structure of 2H-NbSe$_2$, like a Leggett mode or a Bardasis-Schrieffer mode, cannot be easily reconciled with this behavior, since they are intimately related to the properties of the SC state, which barely changes as a function of pressure. For example the Leggett mode, due to the relative fluctuations of the SC phases in two bands, becomes Raman visible thanks to the hole/electron character of the various bands\cite{blumberg_prl07,cea_raman_prb16}, which are not expected to change with pressure.
The Bardasis-Schrieffer mode can manifest as a sharp mode below $2\Delta_{SC}$\cite{monien_prb90}. It originates from subleading pairing fluctuations, so it should be observed in a Raman channel orthogonal to the one where the driving SC instability occurs\cite{monien_prb90}.  Thus, since 2H-NbSe$_2$ is a $s$-wave superconductor the Bardasis-Schrieffer mode should  {\em not} be visible in the $A_{1g}$ channel\cite{monien_prb90}, in contrast with our experimental results.

As we mentioned above, we computed the Raman intensity in the coexisting CDW and SC state with a model Hamiltonian (see Appendix \ref{model})  that is not intended to reproduce realistically 2H-NbSe$_2$. This implies for example that in our approach the CDW originates from a Fermi-surface nesting instability, while in 2H-NbSe$_2$ it has been clearly shown that the electronic susceptibility gets strongly enhanced at the ordering wavevector ${\bf Q}_{CDW}$ only when the momentum dependence of the electron-phonon coupling is taken into account\cite{mauri_prb09,flicker_natcomm15}. However, once this effect is included the description of the CDW state does not differ conceptually from a standard Peierls mechanism\cite{flicker_natcomm15}, making our approach suitable to include the microscopic ingredients specific of 2H-NbSe$_2$.

\section{\label{conclusion}Conclusions}

In summary, we report the pressure dependence of the A$_{1g}$ and E$_{2g}$ Raman active modes related to the charge-density-wave and superconducting orders in the transition metal dichalcogenide 2H-NbSe$_2$ up to $\sim$~5GPa. We showed that the soft CDW modes, the so-called amplitudons, and the sub-gap SC peaks, the Higgs mode, collapse in the pure SC state above 4~GPa while only the expected Cooper-pairs breaking peak at $2\Delta_{SC}$ persists in the E$_{2g}$ symmetry. In the coexisting CDW and SC state, the CDW amplitudon modes soften, enlarge and loose intensity while the intensity of the SC Higgs mode, in both symmetries, decreases but remains sizable even when the amplitudons are almost invisible. These results reveal that the radiance of the SC Higgs mode guarantees that a residual CDW order is present.

At all pressure up to 4~GPa and in both A$_{1g}$ and E$_{2g}$ symmetries, we observed a shift to higher energy of the CDW amplitudons and their enlargement upon entering the SC state as well as a transfer of spectral weight from the CDW amplitudons to the SC Higgs peaks. The pressure trends of the two intertwined CDW and SC modes are well reproduced by our exact calculation of the Raman response within a microscopic model system for the CDW and SC coexisting states. This implies that CDW and SC order parameters must overlap at least in part of the Fermi surface. Thus our experimental and theoretical findings support the Higgs type assignment of the superconducting mode in the coexisting CDW+SC region and in both A$_{1g}$ and E$_{2g}$ channels.

Interestingly, from 0 to 4~GPa, the A$_{1g}$ superconducting Higgs peak softens by 30$\%$ whereas the E$_{2g}$ one is constant. An explanation of this strong symmetry-dependent behavior might requires a calculation within a microscopic model for 2H-NbSe$_2$. Our findings also point out that Raman spectroscopy represents the best suited probe to investigate interplay and competition between charge-density-wave and superconducting coexisting orders, notably when varying the dimensionality, in few-layers systems.

\appendix

\section{Theoretical model}\label{model}
The general derivation of the Raman response in the coexisting CDW+SC state has been recently provided in Ref.\cite{cea_cdw_prb14}.  Its explicit form depends on the band structure and on the electron-phonon coupling. In order to simplify the derivation we adopt here the same model system used in Ref.\ \cite{cea_cdw_prb14}. Even thought it does not provide a complete microscopic description of 2H-NbSe$_2$, it contains the main ingredient needed to describe the interplay of CDW and SC in this system, i.e a momentum-dependent CDW, which leaves part of the Fermi surface ungapped below $T_{CDW}$. This condition makes energetically possible a SC gap opening, and it also allows for a coexisting SC and CDW state in part of the Fermi surface, leading to a coupling between the amplitude fluctuations of the two order parameters.  We
then start from a single-band model on the square lattice (lattice spacing $a=1$)  with band dispersion $\xi_{\bf k}\equiv\epsilon_{\bf k}-\mu=-2t(\cos k_x+\cos k_y)-\mu$,
where $t=1$ is the hopping and $\mu$ is the chemical potential. The CDW instability is driven
by the microscopic coupling $g_{CDW}$ of the electrons to a phonon of energy $\omega_0$
\begin{equation}
\label{hcdw}
H_{CDW}=g_{CDW}\sum_{{\bf k}\sigma} \gamma_{\bf k}
c^\dagger_{{\bf k}+{\bf Q}\sigma}c_{{\bf k}\sigma}(b^+_{\bf Q}+b_{-{\bf Q}}),
\end{equation}
where the $\gamma_{\bf k}=|\cos
k_x-\cos k_y|$ factor modulates the CDW in the momentum space. Near half filling ($\mu=0$)
the nesting of the Fermi surface at the CDW vector ${\bf Q}=(\pi,\pi)$ allows for a CDW instability to occur, with new bands
$\xi_\pm=-\mu\mp\sqrt{\epsilon_{\bf k}^2+\Delta_{CDW}^2\gamma_{\bf k}^2}$ and a CDW order parameter
$\Delta_{CDW}=(4g^2/\omega_0)\sum_{{\bf k}\sigma} \langle \gamma_{\bf k}
c^\dagger_{{\bf k}\sigma} c_{{\bf k}+{\bf Q}\sigma}\rangle$.   The superconductivity originates from a BCS-like interaction term
\begin{equation}
\label{hsc}
H_{SC}=-(U/N)\sum_{q} \Phi_\Delta^\dagger({\bf q})\Phi_\Delta({\bf q}),
\end{equation}
where $\Phi_\Delta({\bf q})\equiv\sum_{\bf k} c_{-{\bf k}+{\bf q}/2\downarrow}c_{{\bf k}+{\bf q}/2\uparrow}$ is
the pairing operator and $N$ is the number of lattice sites. When treated at mean-field level it leads to the
following Green's function $G_0^{-1}(\mathbf{k},i\omega_n)$, defined
on the basis of a generalized 4-components Nambu spinor
$\Psi^\dagger_{\bf k}(i\omega_n)\equiv(c^\dagger_{\mathbf{k}\uparrow}(i\omega_n),
c^\dagger_{\mathbf{k}+\mathbf{Q}\uparrow}(i\omega_n),
c_{-\mathbf{k}\downarrow}(-i\omega_n),
c_{-\mathbf{k}-\mathbf{Q}\downarrow}(-i\omega_n))$ that accounts for
the CDW band folding:
\begin{equation}
\label{G0}
G_0^{-1}(\mathbf{k},i\omega_n)\equiv
i\omega_n-
\begin{pmatrix} \hat h && -\Delta_{SC} \sigma_0\\
-\Delta_{SC} \sigma_0 && -\hat h
\end{pmatrix},
\end{equation}
where $\omega_n=(2n+1)\pi T$ is the fermionic Matsubara frequency, $\Delta_{SC}$ is the superconducting gap, $\sigma_i$ denotes the Pauli matrices and $\hat h$ is a $2\times2$
matrix:
\begin{equation}
\hat h=
\begin{pmatrix} \epsilon_{\bf k}-\mu && -\Delta_{CDW}\gamma_{\bf k}\\
 -\Delta_{CDW}\gamma_{\bf k} && -\epsilon_{\bf k}-\mu
\end{pmatrix}.
\end{equation}
The eigenvalues of the matrix $\hat h$ represent the two CDW bands $\xi_\pm$,
while in the superconducting state the full Green's function (\ref{G0}) has four
possible poles, corresponding to the energies $\pm E_\pm({\bf k})$, with
$E_\pm=\sqrt{\xi_\pm^2+\Delta_{SC}^2}$. For the sake of simplicity we will consider in the following the half-filled case, which allows for an easier treatment of the fluctuations in both superconducting and CDW sectors, without loss of generality of the main conclusions (see Appendix B of ref.\ \cite{cea_cdw_prb14} for further details on the case of general filling).

The Raman response of the previous model has been derived in Ref.\ \cite{cea_cdw_prb14}. Its general structure can be written as
\begin{equation}
\label{rrphonon}
\chi_{RR}(i\Omega_n)=\chi^0_{RR}-g_{CDW}^2\chi^2_{R,CDW}(i\Omega_n)D_{ph}(i\Omega_n)
\end{equation}
where  $\Omega_n=2\pi nT$ is the bosonic Matsubara frequency, $\chi^0_{RR}=\langle \rho_R\rho_R\rangle$ is the bare electronic Raman response, $D_{ph}(i\Omega_n)$ is the Green's function of the ${\bf Q}$ phonon and
$\chi_{R,CDW}=\langle \rho_R \delta \Delta_{CDW}\rangle$ is the response function coupling the electronic Raman density $\rho_R$ to the amplitude fluctuations $\delta \Delta_{CDW}$ of the CDW  order parameter. Eq.\ (\ref{rrphonon}) establishes that when the system enters the CDW state the  phonon coupled via Eq.\ (\ref{hcdw}) to the electronic charge fluctuations at the ${\bf Q}_{CDW}$ ordering vector becomes Raman active. In addition, the spectral function itself of the phonon changes dramatically due to its coupling to the electronic charge fluctuations at ${\bf Q}_{CDW}$. As usual, the coupling of the phonon to the particle-hole excitations is described by  self-energy corrections, which renormalize in general its bare frequency $\omega_0$ and introduce a finite broadening\cite{allen_prb97,Zeyher_Superconductivity_1990,balseiro}:
\begin{equation}
\label{dph}
D_{ph}(i\Omega_n)=-\frac{2\omega_0}{(i\Omega_n)^2-\omega_0^2-\Sigma(i\Omega_n)}.
\end{equation}
However, in contrast to the case considered in \cite{allen_prb97,Zeyher_Superconductivity_1990} of ordinary phonons in metal, the ordering of the electronic charge at ${\bf Q}_{CDW}$  below $T_{CDW}$ implies that the CDW phonon is directly coupled to the amplitude fluctuations of the CDW order parameter\cite{rice_ssc74,browne_prb83,cea_cdw_prb14}, justifying its denomination as "amplitudon". More explicitly one then has
\begin{equation}
\label{sigmaph}
\Sigma(i\Omega_n)=2g^2_{CDW}\omega_0 \chi_{CDW}(i\Omega_n), \quad \chi_{CDW}=\langle \delta \Delta_{CDW} \delta \Delta_{CDW}\rangle.
\end{equation}
After analytical continuation to real frequencies the renormalized phonon frequency $\Omega_0$ below $T_{CDW}$ is then defined, after Eq.\ (\ref{dph}), as a solution of the equation
\begin{equation}
\label{newomega}
\Omega_0^2\equiv \omega_0^2+\Sigma'(\Omega_0)=\omega_0^2[1+(2g_{CDW}^2/\omega_0)\chi'_{CDW}(\Omega_0)]
\end{equation}
 which leads to a temperature dependent $\Omega_0(T)$ scaling approximately as the CDW order parameter. In particular $\Omega_0\rightarrow 0$ as $T\rightarrow  T_{CDW}$ and  $\Omega_0(T=0)$ is much smaller than the bare frequency $\omega_0$\cite{rice_ssc74,cea_cdw_prb14}. As a consequence, near $\Omega_0$ Eq.\ (\ref{rrphonon}) assumes the form of Eq.\ (\ref{chi}), with $\Gamma_{ph}\simeq-\Sigma''(\Omega_0)$  and $Z_{eff}\simeq 2g_{CDW}^2\omega_0\chi'^2_{R,CDW}(\Omega_0)$.

\begin{figure}[h!]
\centering
\includegraphics[width=\linewidth]{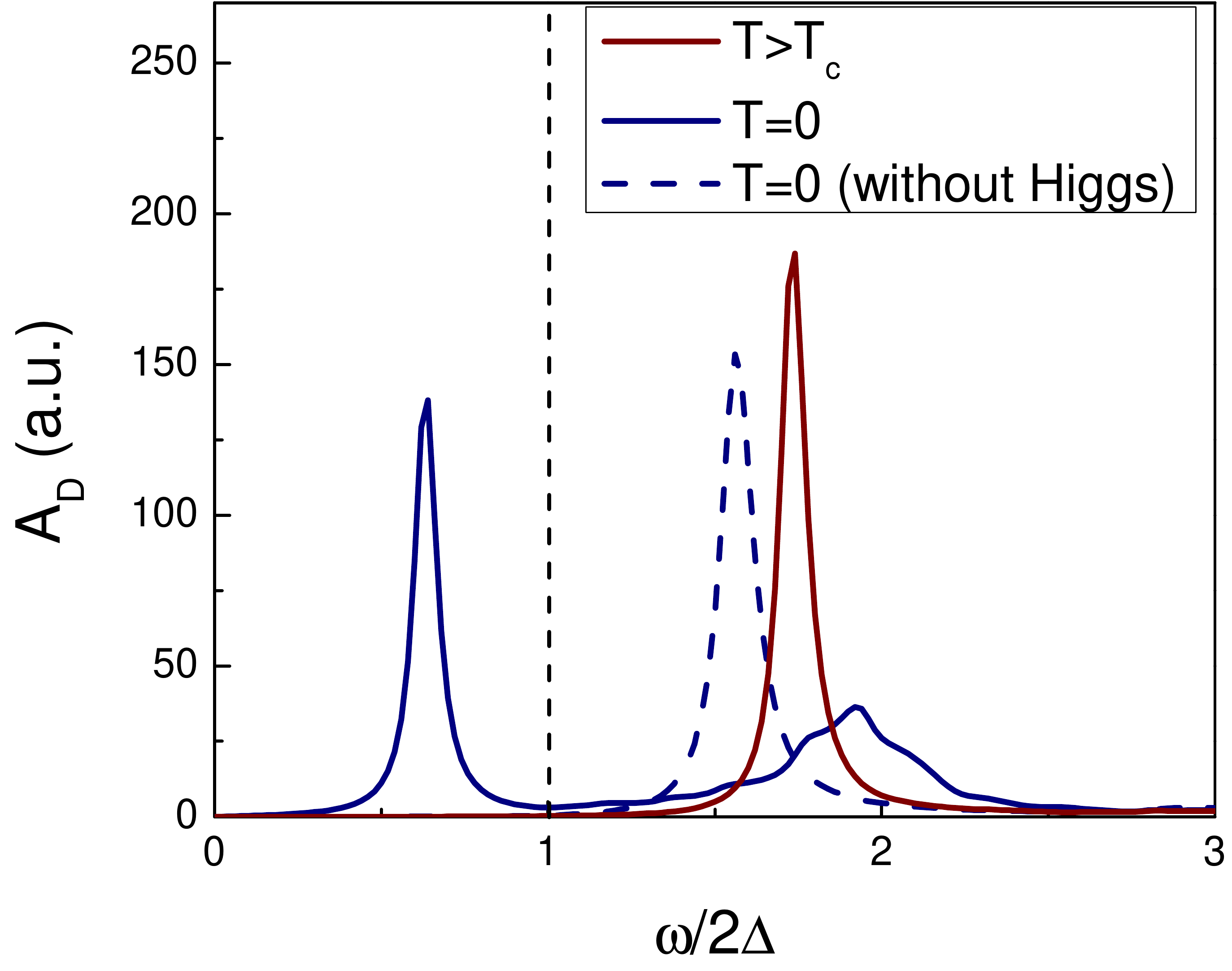}
\caption{Change in the phonon spectral function in the transition from the CDW (solid red line) to the superconducting state with (solid blue line) and without (dashed blue line) coupling to the Higgs. The bare phonon energy is taken at $\omega_0=0.16t=3.2(2\Delta_{SC})$, so the softening of the phonon frequency from $\omega_0$ to a value $\Omega_0$ of the order of $2\Delta_{SC}$ is due to the coupling to the CDW amplitude fluctuations, encoded in self-energy (\ref{sigmaph}). Below $T_c$ the bare self-energy (\ref{sigmaph}) is weakly affected (dashed blue line) by the superconducting gap opening, due to the fact that the electronic excitations were already gapped by the CDW gap. However, the coupling to the Higgs encoded in the full self-energy (\ref{rrsc}) leads to a sharp additional sub-gap peak, and to a hardening and broadening of the phonon spectral function.}
\label{fig:6}
\end{figure}

When entering the superconducting state the phonon self-energy is modified in two ways. First, the two CDW bands $\xi_\pm=\mp\sqrt{\epsilon_{\bf k}^2+\Delta_{CDW}^2\gamma_{\bf k}^2}$ are further gapped by the superconducting gap $\Delta_{SC}$, so that the dispersion becomes $E_\pm=\sqrt{\epsilon_{\bf k}^2+\Delta_{CDW}^2\gamma_{\bf k}^2+\Delta_{SC}^2}$. This reflects in general in a change of the self-energy (\ref{sigmaph}). However, in contrast to the standard case of the metal-to-superconductor transition\cite{allen_prb97,Zeyher_Superconductivity_1990}, the change of quasiparticle dispersion has weak effects on the phonon lineshape, due to the fact that the electronic excitations are already gapped by the CDW gap at $T>T_c$. This is explicitly shown in Fig.\ \ref{fig:6}, where we report the change in the phonon spectral function due only to the modifications of the self-energy (\ref{sigmaph}) below $T_c$. The phonon slightly softens, but no new peak develops below $2\Delta_{SC}$.  However, in the mixed state the phonon self-energy acquires a new  term which represents in diagrammatic language a vertex corrections in the particle-particle channel\cite{varma_prb82,browne_prb83,cea_cdw_prb14}. The full self-energy is then computed as:
\begin{widetext}
\begin{equation}
\label{rrsc}
\Sigma(i\Omega_n)=2g_{CDW}^2\omega_0\chi_{CDW}(i\Omega_n)-2g_{CDW}^2\omega_0\chi^2_{SC,CDW}/X_{SC}
\end{equation}
\end{widetext}

Here $\chi_{SC,CDW}=\langle \delta \Delta_{SC}\delta \Delta_{CDW}\rangle$ is the response function measuring the change in the superconducting amplitude induced by a fluctuation of the CDW gap, and vice versa. As we mentioned in the main text, it provides a direct coupling between the phonon mode and the Higgs mode, whose fluctuations are described by the function $X_{SC}(i\Omega_n)=2/U+\chi_{SC}(i\Omega_n)$, where $\chi_{SC}=\langle \delta \Delta_{SC} \delta \Delta_{SC}\rangle$. More specifically, the Higgs resonance occurs when $X'_{SC}(\omega=2\Delta_{SC})=0$.
 As a consequence, the equation (\ref{newomega}), which defines the poles of the phonon propagator probed by the Raman response (\ref{rrphonon}), acquires an additional solution at $\omega<2\Delta_{SC}$,  see Fig.\ \ref{fig:6}. In addition the phonon signature at $\Omega_0$ changes considerably, with a broadening and hardening analogous to the experimental observations. We also note that,  as   discussed in Ref. \cite{cea_cdw_prb14}, the broadening of the Higgs resonance due to its decay in particle-hole excitations is less pronounced in the coexisting CDW+SC state, since the quasiparticle spectrum above $2\Delta_{SC}$ remains partly gapped by the CDW gap.  This explains why the Higgs resonance coupled to the phonon mode appears so sharp in 2H-NbSe$_2$.

The above Eq.s\ (\ref{rrphonon}), (\ref{sigmaph}) and (\ref{rrsc}) are generic to any band structure, and explain the general mechanism of generation of the amplitudon  below $T_{CDW}$ and of its Higgs signature in the mixed CDW+SC state. To evaluate the pressure dependence of the Raman spectra we computed  explicitly their evolution in our toy model. The various susceptibilities listed above are then easily derived using the definitions of the various operators and of the Green's function (\ref{G0}) above.  We then obtain for the fermionic susceptibility the general structure:
\begin{equation}
\chi_{A}=\sum_{\bf k}
\frac{R_A({\bf k})}
{E_{\bf k}((i\Omega_n)^2-4E_{\bf k}^2)}\tanh(\beta
E_{\bf k}/2)
\end{equation}
where $\beta=1/T$, $E_{\bf k}=\sqrt{\epsilon_{\bf k}^2+\Delta_{CDW}^2 \gamma^2_{\bf k}+\Delta_{SC}^2}$ and the form factor $R_A(\bf k)$ depends on the susceptibility under consideration:
\begin{eqnarray}
\label{rrcdw}
R_{R,CDW}&=&8\Gamma({\bf k})\Delta_{CDW}\gamma^2_{\bf k}\epsilon_{\bf k},\\
R_{CDW}=R_{SC}&=&4\epsilon_{\bf k}^2,\\
R_{SC,CDW}&=&-8\Delta_{SC}\Delta_{CDW}\gamma^2_{\bf k}
\end{eqnarray}

In the present model the Raman response is found different from zero only in the $A_{1g}$ channel where $\Gamma({\bf k})=\cos k_x+\cos k_y\propto \epsilon_{\bf k}$, leading to a term proportional to $\epsilon_{\bf k}^2$ in Eq.\ (\ref{rrcdw}) that survives under momentum integration. For the same reason, we find that the phonon does not couple to the total charge density, as obtained by setting $\Gamma({\bf k})=1$ in Eq.\ (\ref{rrcdw}). This also means that the phonon response is not screened by the long-range Coulomb interactions, mediated by density fluctuations. Even though the present model does not capture the microscopic band structure of 2H-NbSe$_2$, we expect that a similar mechanism is at play in this system as well, explaining the lack of Coulomb screening of the A$_{1g}$ CDW phonon observed experimentally.
The CDW and superconducting order parameters are computed by solving self-consistently the two equations:
\begin{eqnarray}
\label{eqcdw}
\Delta_{CDW}&=&\Delta_{CDW}\frac{2g^2}{\omega_0N}\sum_{\bf k}\frac{\gamma_{\bf k}^2}{E_{\bf k}}\tanh(\beta
E_{\bf k}/2)\\
\Delta_{SC}&=&\Delta_{SC}\frac{U}{2N}\sum_{\bf k}\frac{1}{E_{\bf k}}\tanh(\beta
E_{\bf k}/2)
\end{eqnarray}
Here we performed the calculations for the following choice of parameters at $P=0$: $\omega_0=0.16t$, $g_0=0.14t$, $U_0=0.97t$. To simulate the effect of pressure we suppressed progressively the CDW effective coupling $g^2/\omega_0$ (see Eq.\ (\ref{eqcdw} above) up to the value $g=0.117t$, where CDW order disappears. As a consequence $(g_0^2-g^2)/g_0^2\simeq 2(g-g_0)/g_0=\alpha$ with the definition of the $\alpha=2(g_0-g)/g_0$ given in the text. Simultaneously, we slightly suppressed the SC coupling down to $U=0.79t$, in order to keep $T_c$ almost constant as in the experiments.

Notice that in principle in the present model at half-filling the direct coupling between the Higgs mode and the Raman density, $\chi_{R,SC}=\langle \rho_R \delta \Delta_{SC}\rangle$ is not zero. This is a quite peculiar effect of the band structure considered, that is not expected to hold in 2H-NbSe$_2$ where the bands are approximately parabolic. In this situation indeed the Raman density scales as the total density and the direct coupling of the Higgs mode  to the e.m. field is vanishingly small, as in ordinary superconductors\cite{cea_cdw_prb14,cea_prl15,cea_thg_prb16}. For this reason we did not include explicitly this coupling in the above calculations, and we refer the reader to Ref.\ \cite{cea_cdw_prb14} for a discussion of its role.

\section*{Aknowledgments}

This work was supported by the Labex SEAM (Grant No. ANR-11-IDEX-0005-02), by the French Agence Nationale de la Recherche (ANR PRINCESS, Grant No. ANR-11-BS04-002 and ANR SEO-HiggS2, Grant No. ANR-16-CE30-0014), by the Italian MIUR (PRINRIDEIRON-2012X3YFZ2), by the Italian MAECI under the Italia-India collaborative project SuperTop (PGR04879) and by the Graphene Flagship.  We thank gratefully Chandra Varma for providing insight and motivation, F. Mauri, M. Calandra, P. Rodi\`{e}re, H. Suderow and I. Paul for fruitful discussions and G. Lemarchand and A. Polian for technical support.

%


\end{document}